\documentclass[11pt]{article}

\usepackage{amsmath, amssymb}
\usepackage{algorithmic}
\usepackage{algorithm}
\usepackage{multirow}
\usepackage{bm}
\usepackage{bbm}
\usepackage{graphicx}

\usepackage{booktabs} 
\usepackage{float} 
\usepackage{natbib}

\usepackage{rotating}

\newcommand{\re}{{\rm e}}

\title{Modelling and Bayesian analysis of the Abakaliki Smallpox Data} 

\author{Jessica E. Stockdale, Theodore Kypraios, Philip D. O'Neill\\[2mm] 
School of Mathematical Sciences, University of Nottingham\\ 
\vspace{-5mm}
}
\date{}

\begin{document}

\maketitle 

\begin{abstract}
The celebrated Abakaliki smallpox data have appeared numerous times in the epidemic modelling literature, but in
almost all cases only a specific subset of the data is considered. There is one previous analysis of the full data
set, but this relies on approximation methods to derive a likelihood. The data themselves continue to be of interest
due to concerns about the possible re-emergence of smallpox as a bioterrorism weapon. We present the first full Bayesian
analysis using data-augmentation Markov chain Monte Carlo methods which avoid the need for likelihood approximations.
Results include estimates of basic model parameters as well as reproduction numbers and the likely path of infection.
Model assessment is carried out using simulation-based methods.
\end{abstract}

\section*{Keywords}
Smallpox, Bayesian inference, Markov chain Monte Carlo, Stochastic epidemic model, Abakaliki

\section{Introduction}
In 1967, an outbreak of smallpox occurred in the Nigerian town of Abakaliki. The vast majority of cases were members of the Faith Tabernacle Church (FTC), a religious organisation whose members refused vaccination. A World Health Organization (WHO) report \citep{thompsonfoege68} describes the outbreak, with information on not only the time series of case detections but also their place of dwelling (compound), vaccination status, and FTC membership. The outbreak has inherent historical interest as it occurred during the WHO smallpox eradication programme initiated in 1959. Although smallpox was declared eradicated in 1980, it regained attention as a potential bioterrorism weapon in the early 2000s (see e.g. \cite{G+L}, \cite{Melt} and \cite{Hall}) and continues to be of interest due to concerns about its re-emergence or synthesis (see e.g.  \cite{HenIsao14}, \cite{Eto15}, \cite{WHO15} and references therein). Public health planning for potential future smallpox outbreaks requires estimates of the parameters governing disease transmission, and thus being able
to accurately obtain such quantities from available data is of considerable importance.

Within mathematical infectious disease modelling, the Abakaliki smallpox data set has been frequently cited, the first appearance being \cite{B+T}.
The data are almost always used to illustrate new data analysis methodology, but in virtually all cases most aspects of the data are
ignored apart from the population of 120 FTC individuals and the case detection times, while the models used are not particularly
appropriate for smallpox (see for example \cite{beck}, \cite{Yip}, \cite{ON+R}, \cite{OB}, \cite{Hug}, \cite{B+G}, \cite{L+Y},
\cite{C+O}, \cite{kyp}, \cite{shan}, \cite{X+N}, \cite{McK}, \cite{Gol}, \cite{oh}, \cite{Xu16} and references therein). \cite{RM08} use a more realistic
smallpox model and take account of the compounds where individuals lived, but again ignore all non-FTC individuals.

The main objective of this paper is to present a Bayesian analysis of the full data set.
To our knowledge, the only previous analysis of the full Abakaliki data is in \cite{E+D}, where the authors define a stochastic individual-based transmission model
that considers not only the case detection times but also the other aspects of the data. Their model takes account of the population structure, the disease progression for smallpox,
the vaccination status of individuals and the introduction of control measures during the outbreak. The model parameters are then estimated by constructing and maximising
a likelihood function which is itself constructed using various approximations. Specifically, the true likelihood of the observed data given the model parameters is practically intractable, since it involves integrating over all possible unobserved events, such as the times at which individuals become infected. Eichner and Dietz tackle this problem
by first using a back-calculation method to approximate the distribution of unobserved event times for a given individual, and then by making various assumptions
about independence between individuals in order to construct an approximate likelihood function.

An alternative solution to the intractable likelihood problem is to use data-augmentation methods to produce an analytically tractable (and correct) likelihood, which
can then be incorporated in a Bayesian estimation framework by using Markov chain Monte Carlo (MCMC) methods along the lines described in \cite{ON+R} and \cite{G+R}.
We adopt this approach to carry out a full Bayesian analysis of the Abakaliki smallpox data, whilst also assessing how well
the Eichner and Dietz approximation method works in this setting. Our approach provides results which can be directly compared with those of Eichner and Dietz, specifically
estimates of model parameters, estimates of associated quantities of interest such as reproduction numbers, and the sensitivity of the results to the disease progression
assumptions. In addition, we also estimate quantities derived via data-augmentation, such as who-infected-whom and the time of infection for each individual,
carry out various forms of model assessment to see how well the model fits the data, and explore particular aspects of the model via simulation. None of these
additional elements feature in the Eichner and Dietz analysis.

The paper is structured as follows. In section 2 we describe the data, stochastic transmission model and method of inference. Section 3 contains results
and details of sensitivity analysis and model-checking procedures. We finish with discussion in Section 4. The supplementary material contains details of some
likelihood calculations and the MCMC algorithm.

\section{Data, model and inference methods}\label{sec:pro}
The outbreak is described in detail in \cite{thompsonfoege68} and \cite{E+D}. There were 32 cases in total, 30 of which were FTC members. All of the infected
individuals lived in compounds, which were typically one-storey dwellings built around a central courtyard, and capable of housing several families. The FTC
members frequently visited one another and were somewhat isolated from the rest of the community, which is one reason why most previous data analyses only
consider FTC members. Although FTC members refused vaccination, many of them had been vaccinated prior to joining FTC as described below.

\subsection{Abakaliki smallpox data}
Table \ref{abakaliki} contains details of the 32 cases of smallpox recorded during the outbreak, specifying the date of onset of rash, compound
identifier, FTC membership status and vaccination status. Note that we set a timescale by setting day zero of the outbreak to be the first onset
of rash date. The composition of the affected compounds is provided in Table \ref{compounds}, where the total numbers of vaccinated and non-vaccinated FTC and non-FTC members within each compound are listed. Note that on day 25, four FTC individuals from compound 1 (three vaccinated and one non-vaccinated) moved to compound 2. In addition, quarantine measures were put in place in Abakaliki, but not until part way through the outbreak. The exact time these measures were introduced was not recorded.

\begin{table}[H]
\centering
\begin{tabular}{c|c|c|c|c}
Case number & Day of onset of rash & Compound & Confession & Vaccination\\
\hline
0 & 0 & 1 & FTC & No \\
1 & 13 & 1 & FTC & No \\
2 & 20 & 1 & FTC & No \\
3 & 22 & 1 & FTC & No \\
4 & 25 & 1 & FTC & No \\
5 & 25 & 1 & FTC & No \\
6 & 25 & 1 & FTC & No \\
7 & 26 & 2 & FTC & Yes \\
8 & 30 & 2 & FTC & Yes \\
9 & 35 & 1 & FTC & No \\
10 & 28 & 4 & FTC & No  \\
11 & 40 & 5 & FTC & No  \\
12 & 40 & 1 & FTC & No  \\
13 & 42 & 1 & FTC & No  \\
14 & 42 & 1 & FTC & No  \\
15 & 47 & 1 & FTC & No  \\
16 & 50 & 5 & FTC & No  \\
17 & 51 & 2 & FTC & No  \\
18 & 55 & 1 & FTC & No  \\
19 & 55 & 2 & FTC & No  \\
20 & 56 & 6 & Non & Yes  \\
21 & 56 & 5 & FTC & Yes  \\
22 & 57 & 2 & FTC & Yes  \\
23 & 58 & 7 & FTC & No  \\
24 & 60 & 4 & FTC & No  \\
25 & 60 & 2 & FTC & No  \\
26 & 61 & 2 & FTC & No  \\
27 & 63 & 8 & Non & Yes  \\
28 & 66 & 3 & FTC & No  \\
29 & 66 & 9 & FTC & No  \\
30 & 71 & 5 & FTC & No  \\
31 & 76 & 2 & FTC & Yes\\
\end{tabular}
\caption{Smallpox cases in Abakaliki, Nigeria during 1967, taken from \cite{thompsonfoege68}. Compounds listed are those before the move of cases 7 and 8, and 2 other uninfected individuals, on day 25 from compound 1 to compound 2.}
\label{abakaliki}
\end{table}

\begin{table}
\centering
\begin{tabular}{c|c|c|c|c|c|c}
Compound & \multicolumn{3}{|c|}{FTC} & \multicolumn{3}{c}{Non-FTC} \\
 & Vaccinated & Nonvaccinated & $n_{c,FTC}$ & Vaccinated & Nonvaccinated  & $n_{c,non}$\\
\hline
1 & 18 & 15 & 33 & 0 & 0 & 0\\
2 & 9 & 5 & 14 & 1 & 0 & 1 \\
3 & 2  & 8 & 10 & 0 & 0 & 0 \\
4 & 2 - $i_4$ & 2 + $i_4$ & 4 & 28 + $i_4$ & 1 - $i_4$ & 29 \\
5 & 4 - $i_5$ & 3 + $i_5$ & 7 & 13 + $i_5$ & 2 - $i_5$ & 15 \\
6 & 0 & 0 & 0 & 40 & 3 & 43 \\
7 & 4 - $i_7$ & 1 + $i_7$ & 5 & 12 + $i_7$ & 3 - $i_7$ & 15 \\
8 & 0 & 0 & 0 & 37 & 5 & 42 \\
9 & 0 & 1 & 1 & 26 & 6 & 32 \\
&&&&&&\\
Sum 1-9 & 35 & 39 & 74 & 161 & 16 & 177  \\
Outside & 46 x 35/74 & 46 x 39/74 & 46 & 30903 x 161/177 & 30903 x 16/177 & 30903  \\
\hline
Total &  &  & 120 & & & 31080  \\
\end{tabular}
\newline
\caption{Composition of the compounds affected by smallpox in Abakaliki, Nigeria during 1967, as described in \cite{thompsonfoege68}.
Since we do not have complete vaccination status for all compounds, we use $i_4,i_5,i_7$ to allow for different possible configurations. The total number of vaccinated individuals is known and so $i_4+i_5+i_7=4$. It is also known that $i_4 \in \{0,1\}, i_5 \in \{0,1,2\} $ and $ i_7 \in\{1,2,3\}$. Note: this table displays the compound composition after the move of the 4 individuals from compound 1 to compound 2 on day 25.}
\label{compounds}
\end{table}

\subsection{Stochastic transmission model}
We suppose that the residents of Abakaliki form a closed population with $N=31,200$ individuals, labelled $0, 1, \ldots, N-1$. Individuals $0, 1, \ldots, n_{com}-1$ are those inside the compounds, where $n_{com}=251$ is the number of people within the compounds. Any individual $k=0, \ldots, N-1$ may be categorised as type $(c_k,f_k)$, where (i) $c_k=1, \ldots, 9$ is the compound of $k$, with $c_k=0$ if $k$ is outside the compounds, and (ii) $f_k$ is $k$'s confession; FTC or non-FTC. These types may lead to differences in the mixing behaviour of individuals, but otherwise individuals are considered to be identical in their susceptibility to smallpox and their ability to infect others.

We now describe a stochastic disease-transmission model for the spread of smallpox throughout the population of Abakaliki. This model is essentially the same as that described in \cite{E+D}, and is a variant of an SEIR (Susceptible-Exposed-Infective-Removed) model. At any given time $t$ each individual in the population will be in any one of six states, namely susceptible, exposed, with fever, with rash, quarantined or removed. For $j=0, \ldots, N-1$, let $e_j$, $i_j$, $r_j$, $q_j$, $\tau_j$ denote, respectively, the times of exposure, fever, rash, quarantine and recovery for individual $j$. Any susceptible individual may become exposed, as described below, at which point they enter an incubation (or latent) period. They next enter the fever stage of the disease, at which point they become infectious and may hence infect others. During the rash stage which follows, the individual remains infectious but with a potentially different level of infectivity. We define the infectious period to be the combined time spent in the fever and rash stages. Infectious individuals will either become removed (namely recovery or death; we do not distinguish these) or isolated, in which the individual is quarantined and henceforth unable to infect others. Control measures, in which cases are placed into isolation soon after detection, are introduced part way through the outbreak at time $t_q$. We do not allow re-infection, so that individuals who have been infected cannot become susceptible again. The epidemic continues until there are no infectious or exposed individuals remaining in the population, at which point each person will either still be susceptible, or will have been quarantined/removed.

The lengths of time spent in each stage of the disease for different individuals are assumed to be mutually independent random variables with specified distributions, the parameter
values of which are assumed known. We adopt the assumptions of the Eichner and Dietz model, so that the incubation period, fever period and rash period all have Gamma distributions
with values as shown in Table \ref{distributions}.
If quarantine measures have been introduced, then an individual may be put into isolation after a random delay following their rash onset date. Specifically, we define the quarantine
time of individual $j$ as $q_j = \text{max}(r_j,t_q)+\Gamma(2,2)$, where $\Gamma(\mu,\sigma)$ denotes a gamma distributed random variable with mean $\mu$ and standard deviation $\sigma$.
This means that no quarantining occurs prior to time $t_q$, after which it takes an average of two days for a detected individual to be placed in isolation. Note that both
removal and quarantining of an individual are equivalent in the sense that both mean the individual can no longer infect others, but we include both in the model for clarity, and
also for comparison with the Eichner and Dietz model. Note also that an infected individual will have both a removal time and a quarantine time, and both quantities appear in the
likelihood function as explained later.

\begin{table}
\centering
\begin{tabular}{c|c|c}
& \parbox[t]{0.6in}{\centering Mean \par (days) \strut} & \parbox[t]{0.7in}{\centering Standard \par  deviation (days) \strut}  \\
\hline
Period before fever & $\mu_I=11.6$ & $\sigma_I=1.9$  \\
From fever to rash & $\mu_F=2.49$ & $\sigma_F=0.88$ \\
From rash until recovery & $\mu_R=16.0$ & $\sigma_R=2.83$  \\
 \begin{tabular}{@{}c@{}}From rash to quarantine \\ or from $t_q$ to quarantine\end{tabular}  & $\mu_Q=2.0$ & $\sigma_Q=2.00$  \\
\end{tabular}
\caption{Durations of disease stages in the smallpox model. The time until quarantine changes after the introduction of control measures as described in the text. }
\label{distributions}
\end{table}

We assume that the epidemic is initiated by a single exposed individual, whom we label $\kappa$. The epidemic thus begins at time $e_{\kappa}$ with the exposure of the initial infective $\kappa$. Recall that the infectious period is defined in two parts: the fever period and the rash period, during each of which an individual will be infectious, but at potentially different rates. During their rash period, an individual $j$ will have contacts with other members of their compound who are of the same confession at times given by the points of a Poisson process of rate $\lambda_h$ per day. Individuals outside of the nine compounds do not have such contacts. In addition, FTC individuals will have contacts at rate $\lambda_f$ per day with other FTC individuals and contacts at rate $\lambda_a$ per day with anybody (including FTC individuals) in the population. Non-FTC individuals are assumed to have contacts with anybody in the population at rate $\lambda_a + \lambda_f$ per day. This assumption is made to ensure that all individuals have the same average number of contacts per day outside of
their own compound. During the fever period, contacts occur in exactly the same manner except that all rates are multiplied by a factor $b$, to account for a potential difference in infectivity during the fever period. In each case, the individual actually contacted is chosen uniformly at random from the pool of potential individuals in question. For example, contacts made by an individual with the entire population are drawn from the $N-1$ other individuals. Note that this means that the individual-to-individual contact rate for such contacts is $\lambda_a/(N-1)$. Any contact from an infective to a susceptible results in immediate exposure of the susceptible. All of the Poisson processes describing contacts are assumed to be mutually independent.

In addition, a proportion of the population is vaccinated. The vaccination status of all but a few individuals within the compounds is known, and the proportions of FTC/non-FTC vaccinated individuals outside the compounds is assumed to be the same as inside. However, vaccination is not necessarily effective: each recipient of the vaccine is  completely protected with probability $v$, or remains completely susceptible with probability $1-v$. Although the total number of vaccinated individuals is known, we do not have complete information on the composition of individuals with respect to vaccination status and FTC membership (see Table \ref{compounds}). There are five potential configurations of twelve individuals with unknown details to consider. For each individual in the population we will have a vaccination status, which is assumed known for most individuals, and a protection status, which is unknown.

\subsection{Reproduction numbers}
Within mathematical epidemic modelling, the so-called basic reproduction number is of primary importance. It can be defined as the average number of cases caused by
a typical index case in a large population, and its value typically governs certain aspects of the epidemic such as the final number of cases or the probability of an epidemic
dying out rapidly. From a mathematical viewpoint, reproduction numbers for stochastic epidemics are typically defined by allowing the population size to become infinite in
an appropriate sense. For models featuring structured populations with different kinds of contact rates, reproduction numbers can be defined in various ways.
Eichner and Dietz consider two such reproduction numbers, namely $R_0 = (\mu_R+b\mu_F)(\lambda_a + \lambda_f + \lambda_h)$ and $R_F = b\mu_F(\lambda_a + \lambda_f + \lambda_h)$. Here, $R_0$ is a reproduction number for an infected FTC member within the compounds, and can be interpreted as the average number of new infections such an individual would cause, under the assumption that the FTC population, compound populations and entire population are all large. Similarly, $R_F$ describes the average number of new infections caused by contacts made during the fever period of an FTC individual within the compounds. As discussed later, one drawback with such definitions for the Abakaliki data is that the compounds themselves
are not particularly large. However, for comparison purposes we shall also consider $R_0$ and $R_F$ in our analysis.

\subsection{Bayesian inference and MCMC}
Our aim is to perform Bayesian inference for the unknown model parameters given the data, which consist of rash times for all infectives, knowledge of the population structure and vaccination status of individuals. We will use an MCMC algorithm to obtain approximate samples from the posterior density of the model parameters, namely the infection rates $\lambda_a$, $\lambda_f$ and $\lambda_h$, the vaccine efficacy $v$, the infectivity factor $b$ and the time quarantine measures were introduced, $t_q$. Our approach involves data augmentation,
specifically involving the exposure, fever, removal and quarantine times of each infected individual, and also the protection statuses and unknown vaccination statuses.
First we derive an expression for the likelihood of the observed and augmented data. Let
\begin{equation*}
\mathbf{\Phi}=(\kappa, e_{\kappa}, t_q, b, v, \lambda_a, \lambda_f, \lambda_h,\mbox{\boldmath{$\theta$}},\mathbf{s})
\end{equation*}
where
\begin{equation*}
\mbox{\boldmath{$\theta$}}=(\mu_I,\sigma_I,\mu_F,\sigma_F,\mu_R,\sigma_R,\mu_Q,\sigma_Q),
\end{equation*}
so that the components of $\mbox{\boldmath{$\theta$}}$ are parameters that are assumed known, and where $\mathbf{s}=(s_0,s_1,...,s_{N-1})$, where  for $i=0, 1, \ldots , N-1$, ${s_i}$ is equal to 1 if individual $i$ is vaccinated and $0$ if not. These vaccination statuses are assumed known for the majority of individuals, with a small number of exceptions, as shown in Table \ref{compounds}.

We define $\mathbf{e}$, $\mathbf{i}$, $\mathbf{q}$ and \mbox{\boldmath{$\tau$}} as the unknown sets of exposure (not including the initial exposure $e_{\kappa}$), infection, quarantine and removal times, respectively. Similarly, we define $\mathbf{r}$ as the known set of rash times for all infectives. Then we define the augmented data as
\begin{equation*}
\mbox{\boldmath{$\gamma$}}=(\mathbf{e},\mathbf{i},\mathbf{q},\mbox{\boldmath{$\tau$}},\mathbf{p}),
\end{equation*}
where $\mathbf{p}=(p_0,p_1,...,p_{N-1})$ contains the unknown protection status of each individual, specifically with $p_i=1$ if individual $i$ is protected, and $p_i=0$ if they are not.

For an individual $j=0,...,N-1$ who is susceptible at time $t$ we define $\Lambda_j(t)$ as the infectious pressure acting upon them at time $t$, so
that
\[
\mathbb{P}(j \text{ is exposed in } (t,t+\delta t] ) = \Lambda_j(t) \delta t +o(\delta t),
\]
whilst for an individual $j$ who is no longer susceptible at time $t$ we set $\Lambda_j(t) = 0$.
From the model definition, if $j$ is susceptible at time $t$ then $\Lambda_j(t)$ can be written as
\begin{equation*}
\Lambda_j(t) =  \sum\limits_{k \in \mathcal{N}_{inf}(t)} m(k,t)  \times
 \begin{cases}
      \frac{\lambda_a}{N-1} + \frac{\lambda_f\delta_f(k,j)}{n-1} + \frac{\lambda_h\delta_c(k,j;t)}{n_{c,f_k}(t)-1} & \text{if } f_k=\text{FTC, }  \\
      \frac{\lambda_a+\lambda_f}{N-1} + \frac{\lambda_h\delta_c(k,j;t)}{n_{c,f_k}(t)-1} & \text{otherwise,} \\
   \end{cases}
\end{equation*}
where
\begin{equation*}
m(k,t) =
 \begin{cases}
     b & \text{if } i_k\leq t<r_k, \\
     1 & \text{if } r_k\leq t<\text{min}(\tau_k,q_k), \\
     0 & \text{otherwise,} \\
   \end{cases}
\end{equation*}
and (i) $\delta_f(j,k)$ = 1 if both $k$ and $j$ are FTC and zero otherwise; (ii) $\delta_c(j,k;t)$ = 1 if both $k$ and $j$ live in the same compound at time $t$ and are of the same confession, and zero otherwise; (iii) $n=120$, the number of FTC individuals in the population; (iv) $n_{c,f_j}(t)$ is the number of individuals in $j$'s compound of the same confession as $j$ at time $t$, including $j$ themselves, and (v) $\mathcal{N}_{inf}(t)$ is the set of individuals infective at time $t$.

We denote the likelihood of the data $\mathbf{r}$ given the model parameters $\mathbf{\Phi}$ as $\pi(\mathbf{r} | \mathbf{\Phi})$. This is practically intractable since
its evaluation involves integrating over all possible unobserved events. We instead proceed by augmenting
the data $\mathbf{r}$ with $\mbox{\boldmath{$\gamma$}}$ to obtain the tractable likelihood
\begin{eqnarray}\label{eq:like}
\pi(\mathbf{r},\mbox{\boldmath{${\gamma}$}} | \mathbf{\Phi}) & = &
\left( \prod_{j \in \mathcal{N}_{inf}} \Lambda_j(e_j-)\right) \times \re^{-\int_{e_{\kappa}}^T \Lambda(t) dt} \nonumber \\
& & \times \prod_{j \in \mathcal{N}_{inf}} f_I(i_j-e_j)f_F(r_j-i_j)  f_R(\tau_j-r_j)f_Q(q_j- \max(r_j,t_q))  \nonumber \\
&& \times v^{\sum\limits_{r=0}^{N-1}p_r s_r }(1-v)^{\sum\limits_{r=0}^{N-1}(1-p_r) s_r },
\end{eqnarray}
where (i) for $t \geq e_{\kappa}$,
\[
\Lambda(t) = \sum_{j=0}^{N-1} \Lambda_j(t)
\]
denotes the total pressure acting on all susceptible individuals at time $t$;
(ii) $\Lambda_j(e_j-) = \lim_{t \uparrow e_j} \Lambda_j(t)$ is the pressure on $j$ just before their exposure; (iii) $\mathcal{N}_{inf}$ is the set of individuals who
ever become infected; (iv) $T$ is the end of the epidemic, i.e. the first time at which no infectives or exposed individuals remain in the population (in practice we set $T$ equal to the final rash time); and (v) $f_A$, for $A=(I$, $F$, $R$, $Q$), is the probability density function of a $\Gamma(\mu_A, \sigma_A)$ distribution. The augmented likelihood function in (\ref{eq:like}) is of a fairly standard form (see e.g. \cite{OB}) and contains the following components. The first product term accounts for the exposure of each of the observed cases and the exponential term gives the probability of individuals avoiding infection (either until they become infected, or throughout the entire epidemic). The second product term gives the likelihood of the times spent in each of the disease progression states for each individual who ever becomes infected. The final terms give the probability of the protection statuses for all individuals in the population.

One practical drawback with our data augmentation scheme as it stands is that it includes protection statuses $p_i$ for all $N = 31,200$ individuals in the population. However, it is possible
to integrate out these parameters for all individuals outside the compounds, essentially because the number of protected individuals follows a Binomial distribution. The calculations are fairly lengthy and so are provided in the supplementary material.

By Bayes' Theorem, the posterior density of interest is
\[
\pi(\mathbf{\Phi}, \mbox{\boldmath{$\gamma$}} | \mathbf{r}) \propto \pi(\mathbf{r},\mbox{\boldmath{${\gamma}$}} | \mathbf{\Phi}) \pi(\mathbf{\Phi}),
\]
where $\pi(\mathbf{\Phi})$ denotes the prior density of $\mathbf{\Phi}$.
We assume {\em a priori} independence of the components of $\mathbf{\Phi}$ so that
\[
\pi(\mathbf{\Phi})  =  \pi(\kappa)\pi(t_q)\pi(b)\pi(v) \pi(\lambda_a)\pi(\lambda_f)\pi(\lambda_h)\pi(\mbox{\boldmath{$\theta$}}) \pi(\mathbf{s}).
\]
We set $\lambda_a$, $\lambda_f$ and $\lambda_h$ to have $\Gamma(10^3, 10^6)$ prior distributions which corresponds to vague prior assumptions for these parameters, set $v$, $b$ and $t_q$ to have uniform prior distributions on $(0,1)$, $(0,\infty)$ and $(0,\infty)$ respectively, and set $\kappa$ to have a discrete uniform distribution over all infected individuals. Since $\mbox{\boldmath{$\theta$}}$ is assumed known, $\pi(\mbox{\boldmath{$\theta$}})$ is just a point mass. Finally, $\pi(\mathbf{s})$ consists of a point mass at the known vaccination statuses with a uniform distribution over the five possible configurations of twelve unknown vaccination statuses as shown in Table \ref{compounds}.

We use an MCMC algorithm to produce samples of the parameters of interest from the target posterior distribution, updating both the model parameters and the unknown event times as well as protection statuses and vaccination combinations. The algorithm is non-standard, and although it is similar in principle to that in \cite{ON+R}, in practice it is far more complex
and involves considerable book-keeping. Full details of the algorithm are provided in the supplementary material.

\section{Results and analysis}
\subsection{Parameter estimates and reproduction numbers}
\begin{figure}
\centering
\includegraphics[scale=0.45]{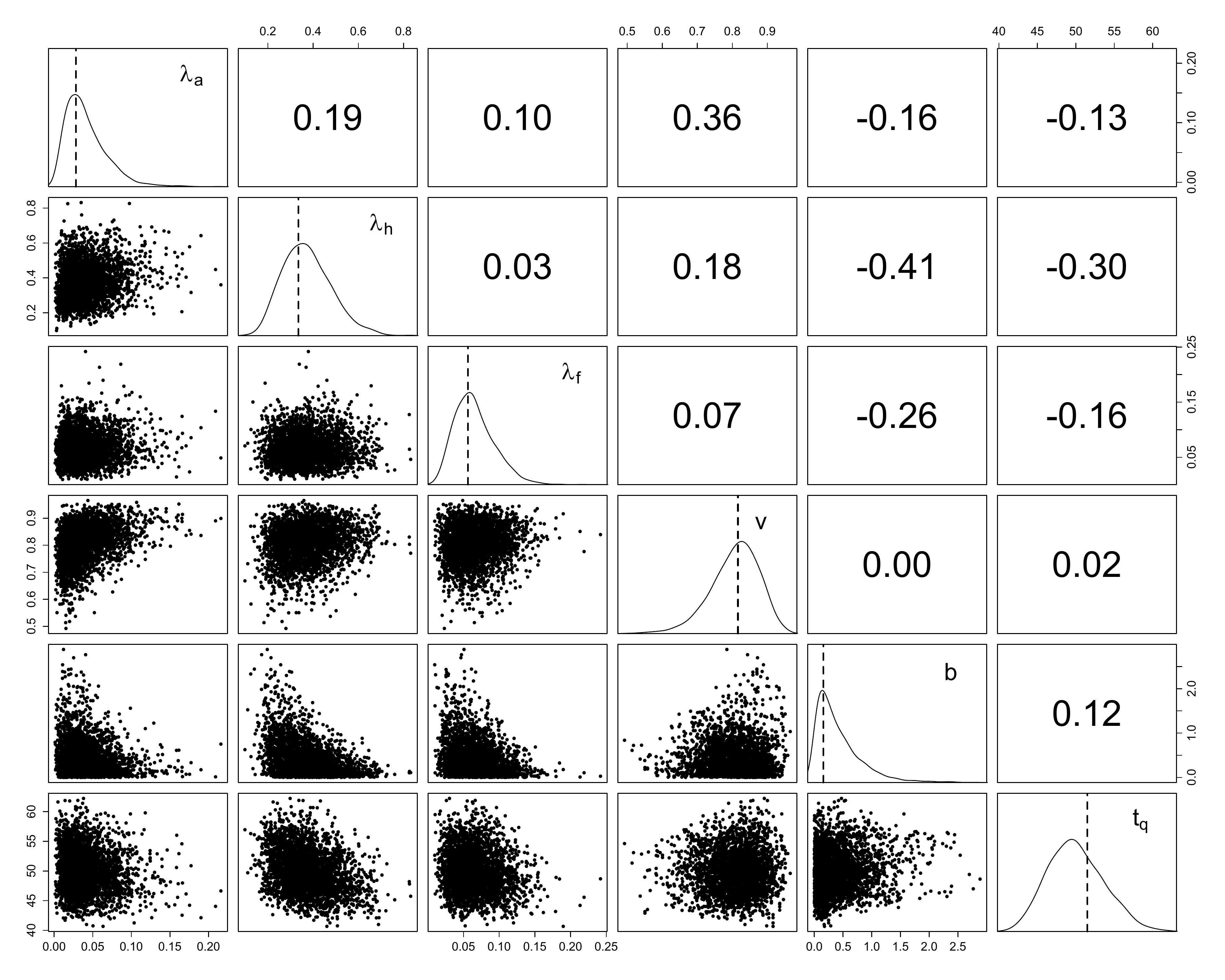}
\caption{The diagonal shows posterior densities for the 6 parameters of interest, based on 100,000 MCMC samples. Dotted lines give the Eichner and Dietz estimates. The lower panels
show scatter plots of the MCMC samples and the upper panels show posterior correlation coefficients.}
\label{fig:density}
\end{figure}

Table \ref{results} contains posterior summaries for the model parameters along with the corresponding maximum likelihood estimates from Eichner and Dietz' approximate likelihood
method, and Figure \ref{fig:density} contains corresponding density plots, scatter plots and posterior correlation coefficients. Our estimates are fairly similar to those of Eichner and Dietz, and in particular the posterior modal values for the six basic model parameters are quite close. Although they represent different quantities, our posterior credible intervals and the confidence intervals of Eichner and Dietz are also quite similar. Our mean estimate of the basic reproduction number $R_0$ is 7.96, which is slightly higher than Eichner and Dietz' estimate of 6.87. Similarly, our estimate of the reproduction number for the fever period $R_F$ is 0.53 compared to Eichner and Dietz's 0.164. In this case the difference can be explained by the highly skewed posterior density for $b$, so that the mean and mode are clearly different. The scatter plots and correlation coefficients suggest that the basic model parameters can
be separately estimated from the data and that the model is not over-parameterised.

\begin{table}
  \centering
\begin{tabular}{l|l|l|l|l|l}
Parameter & Posterior & Posterior  & Credible &   ED  &  ED  confidence \\
 & mean & median & interval & estimate & interval\\
\hline
$\lambda_a$ & 0.041 & 0.035 & (0, 0.093) & 0.0281 & (0.00447, 0.101)  \\
$\lambda_f$ & 0.063 & 0.059 & (0.009, 0.010) & 0.0562 & (0.0187, 0.127) \\
$\lambda_h$ & 0.358 & 0.349 & (0.150, 0.565) & 0.335 & (0.192, 0.527) \\
$v$ & 0.808 & 0.817 & (0.668, 0.947) & 0.816 & (0.644, 0.922)  \\
$b$ & 0.522 & 0.374 & (0, 1.500) & 0.157 & (0, 1.89) \\
$t_q$ & 50.4 & 50.2 & (42.4, 58.4) & 51.5 & (44.7, 59.6) \\
$R_0$ & 7.96 & 7.79 & (4.33, 11.56) & 6.87 & (4.52, 10.1) \\
$R_F$ & 0.531 & 0.431 & (0, 1.364) & 0.164 & (0, 1.31)
\end{tabular}
\caption{Parameter estimates and equal-tailed 95\% posterior credible intervals for the Abakaliki smallpox outbreak from the true likelihood approach, with results from \cite{E+D} for comparison (ED). 100,000 MCMC samples were used. Here $R_0 = (\mu_R+b\mu_F)(\lambda_a + \lambda_f + \lambda_h)$ and $R_F = b\mu_F(\lambda_a + \lambda_f + \lambda_h)$.}\label{results}
\end{table}

\subsection{Who infected whom}
\begin{figure}
\centering
\includegraphics[width=0.75\linewidth]{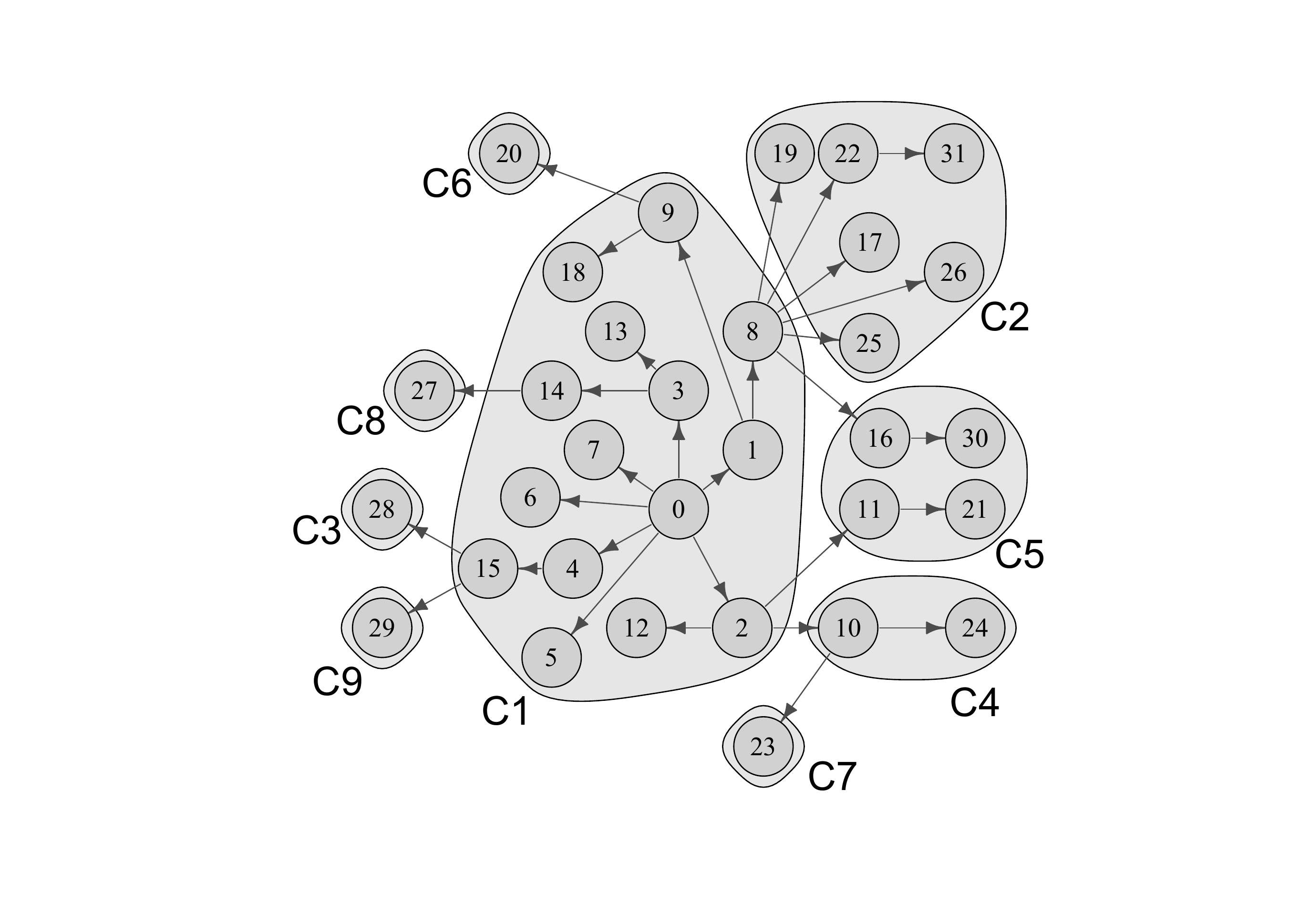}
\includegraphics[width=0.75\linewidth]{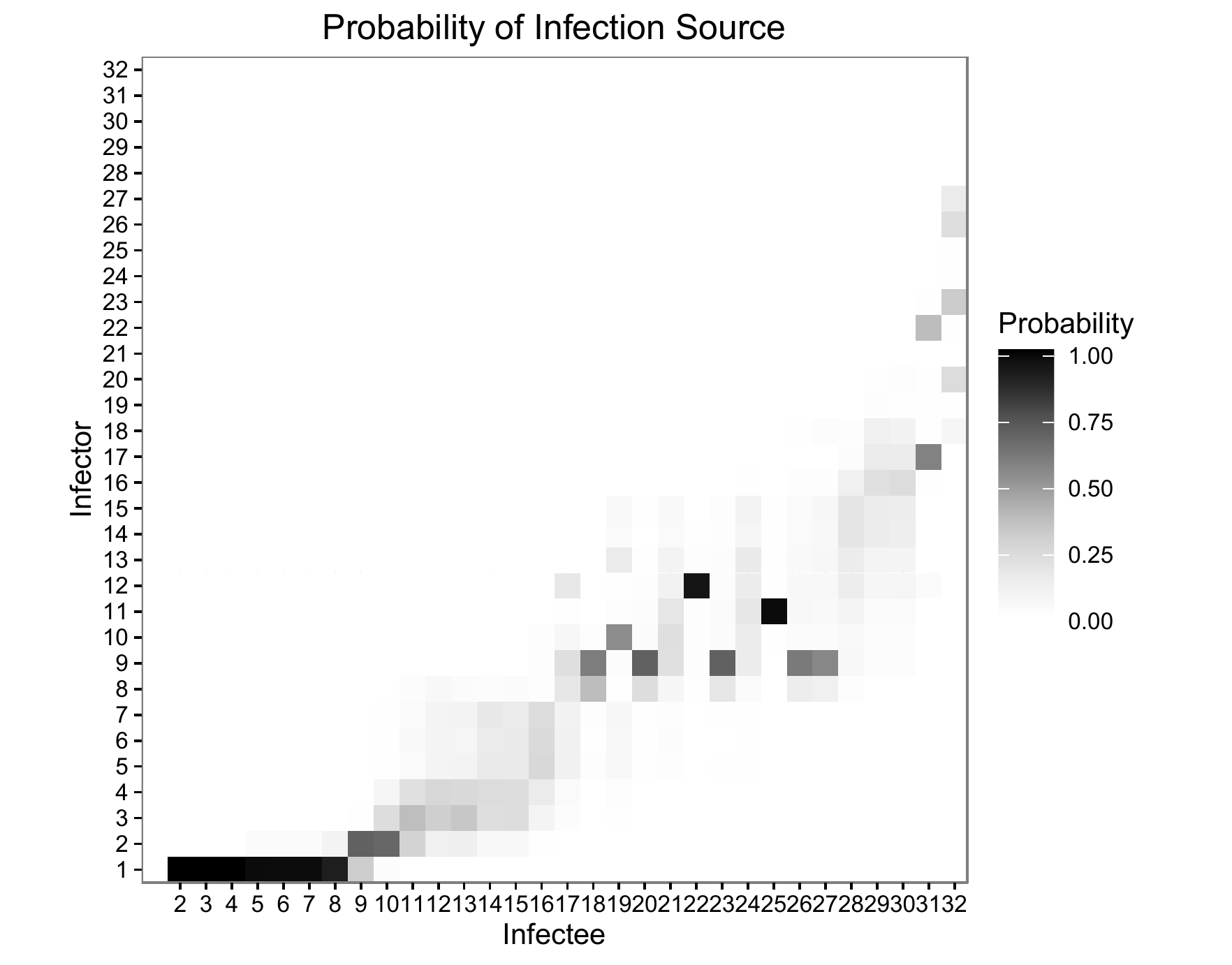}
\caption{Estimation for who-infected-whom during the Abakaliki outbreak. The first figure shows the most likely infector for each individual, i.e. the infector shown has the
highest posterior probability among all possible infectors. Labels C1, ..., C9 denote compounds 1, ..., 9, respectively. Individuals 7 and 8 moved into compound 2 during the outbreak; the figure shows the initial configuration. The second figure shows the posterior distribution of possible infectors for each infectee.}
\label{whoinfwhom}
\end{figure}

Since the MCMC algorithm involves imputation of all event times, it is straightforward to obtain estimates of the path of infection, i.e. who infected whom. Specifically,
if an individual $j$ is subject to infectious pressure $\Lambda_j(t) = \sum_{k=1}^m a_k(t)$ at the time of their exposure, where $a_k(t)$ is the pressure from the $k$th of
$m$ infectives at time $t$, then the probability that individual $k$ actually infected $j$ is simply $a_k(t) / \Lambda_j(t)$. Figure \ref{whoinfwhom} shows the most likely infector
for each observed case and also a greyscale plot which illustrates the associated uncertainty. We see that most infections occurred within compounds; note that individuals 7 and 8 moved from compound 1 to compound 2 during the outbreak and so in reality most of the infections caused by individual 8 were probably also within-compound. Most individuals give rise to one
secondary case, but individuals 0 and 8 both cause multiple secondary cases. The greyscale plot shows that there is a modest degree of uncertainty around the identity of each infector.

\subsection{Exposure times}
\begin{figure}
\centering
\includegraphics[width=0.75\linewidth]{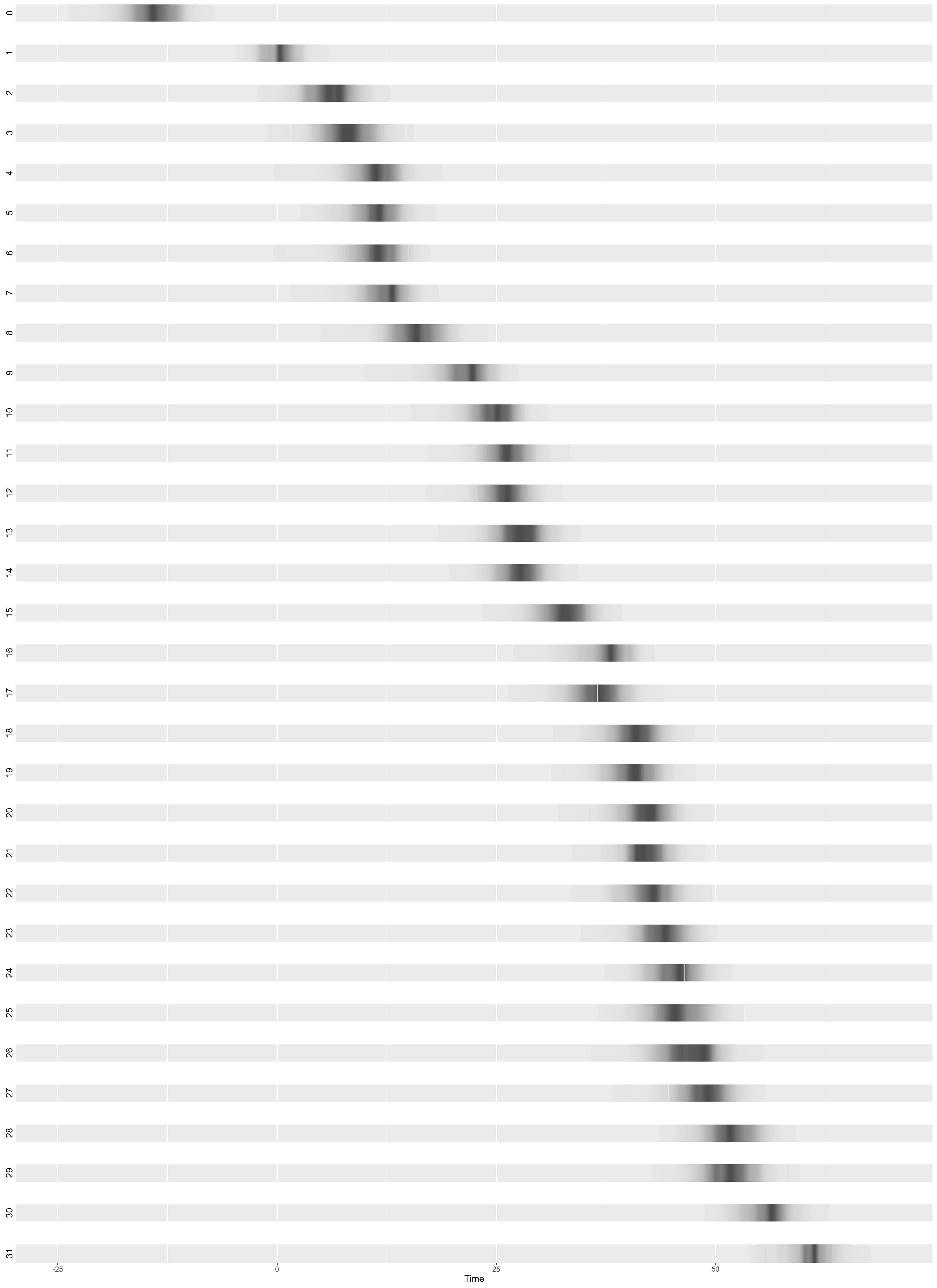}
\caption{Heatmap showing the posterior distribution of time of exposure for each of the 32 cases in the Abakaliki outbreak.}
\label{exposure}
\end{figure}

Figure \ref{exposure} illustrates the posterior distribution of the initial exposure time for each of the 32 cases. Generally speaking there is relatively little uncertainty,
and most of the exposure times follow the ordering of the rash times, both features that are likely to be consequences of the distributions assumed for disease progression.
The figure also shows the temporal aspects of the outbreak in terms of generations of infection: the first two generations (i.e. those infected by the index case, and those
they infect) are clearly discernible, while the third and fourth generations are less distinct from each other, although (according to Figure \ref{whoinfwhom}) the fourth generation
only contains two individuals. We see some groups of individuals with very similar exposure times and who are all infected by the same person, according to Figure \ref{whoinfwhom},
individuals 4, 5, 6 and 10, 11, 12 being two examples. Such clustering, more akin to a point-source outbreak, illustrates the high transmission potential for smallpox in close-contact
settings. Finally, we comment that \cite{E+D} also provide a plot showing likely exposure times, along with other event times, but that this is based entirely on back-calculation using
the assumed disease progression model. In particular, their plot takes no account of the transmission model itself.

\subsection{Sensitivity analysis}
\begin{figure}
\centering
\includegraphics[width =0.75\linewidth]{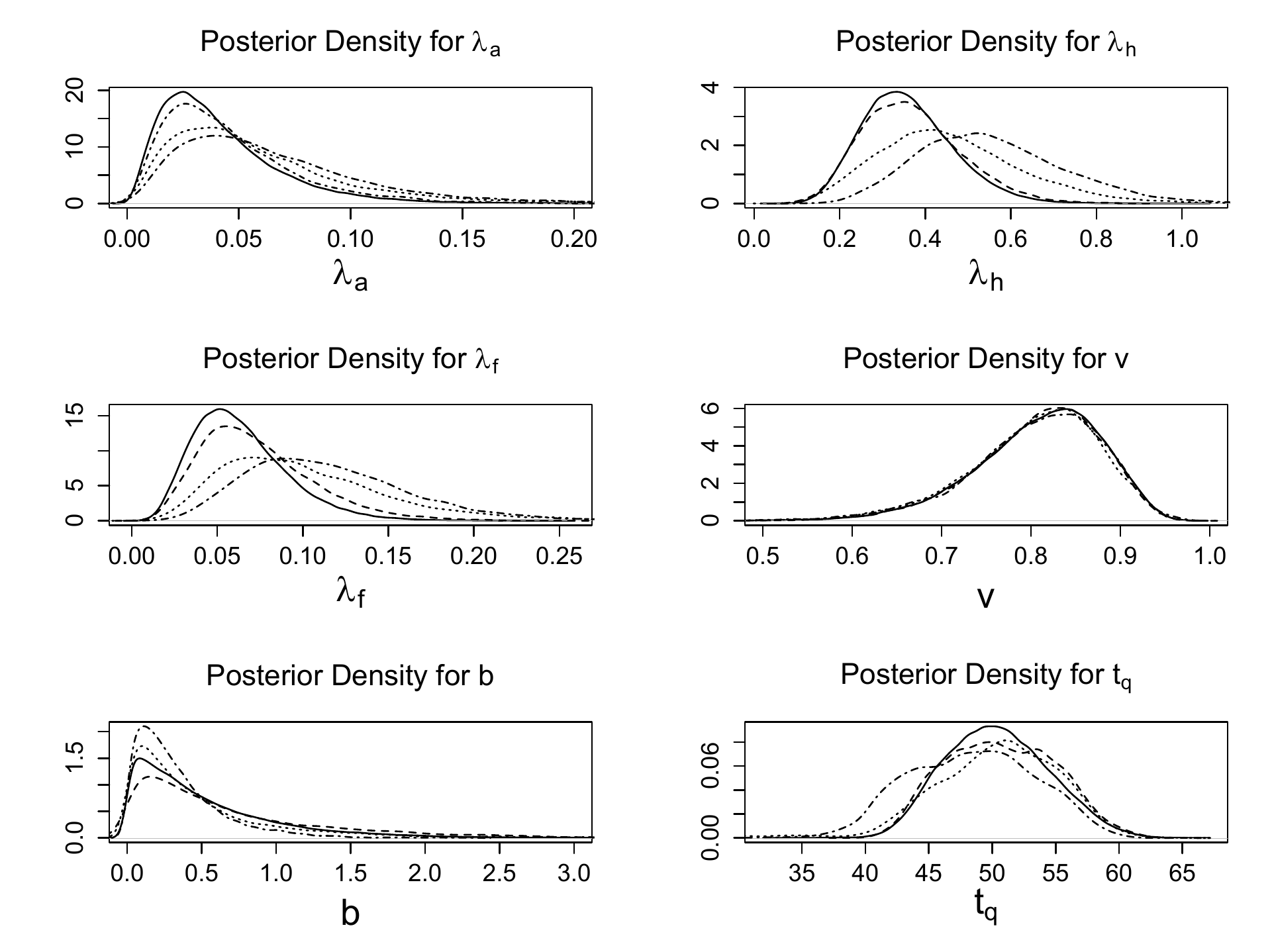}
\includegraphics[width =0.75\linewidth]{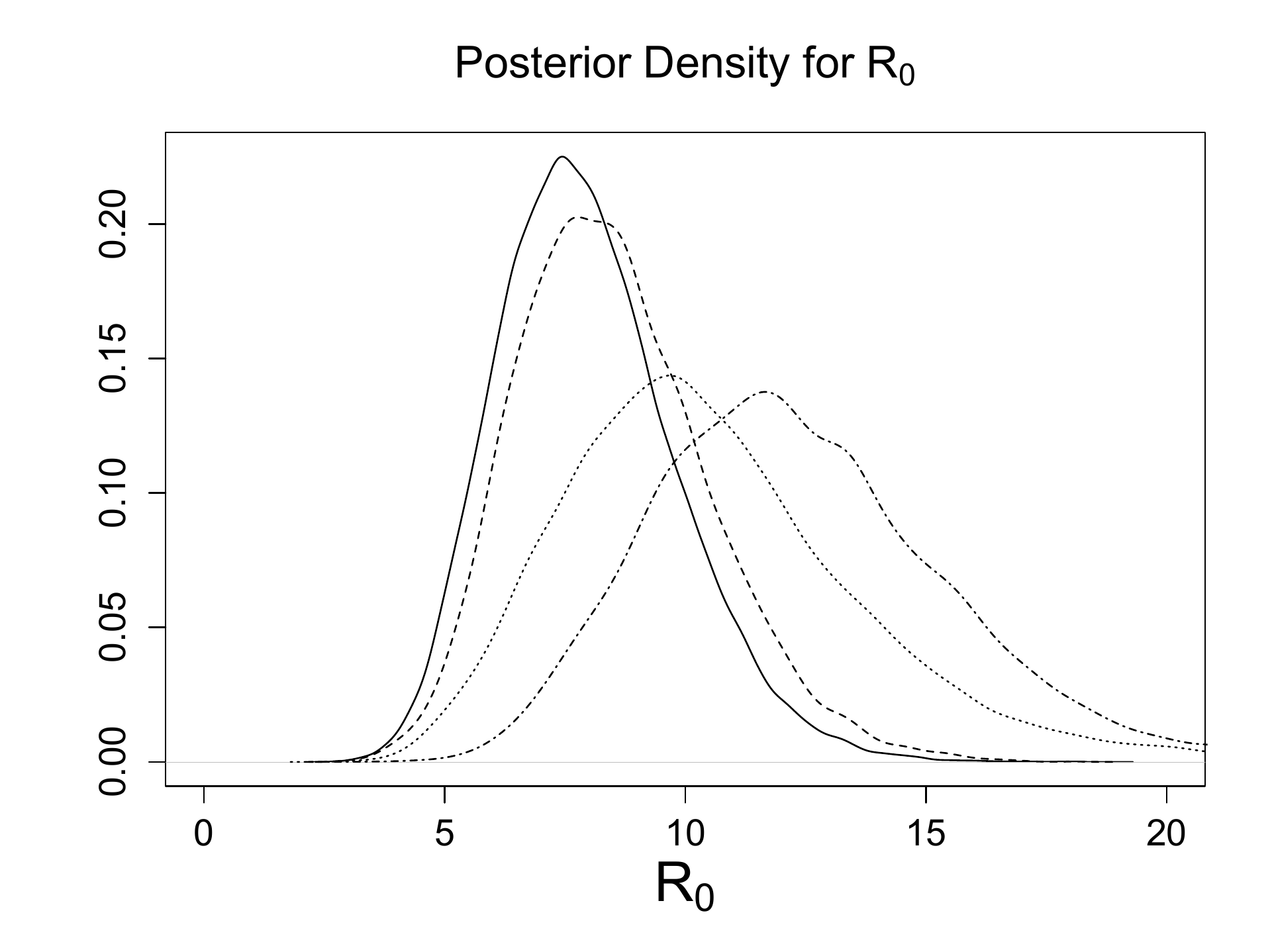}
\includegraphics[scale=0.5, trim={7cm 6cm 4cm 6cm},clip]{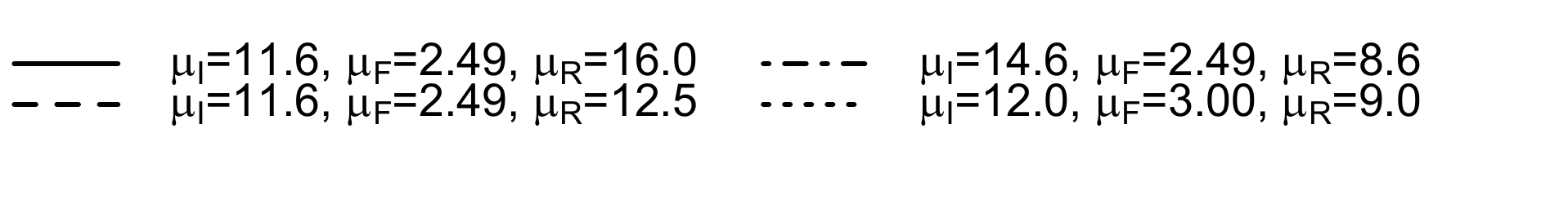}
\caption{Posterior densities of the model parameters and $R_0$ using different mean durations of the disease stages. The solid line represents the baseline case as used in Eichner and Dietz and our primary analysis.}
\label{fig:sdensity}
\end{figure}

\begin{figure}[H]
\centering
\includegraphics[width =0.75\linewidth]{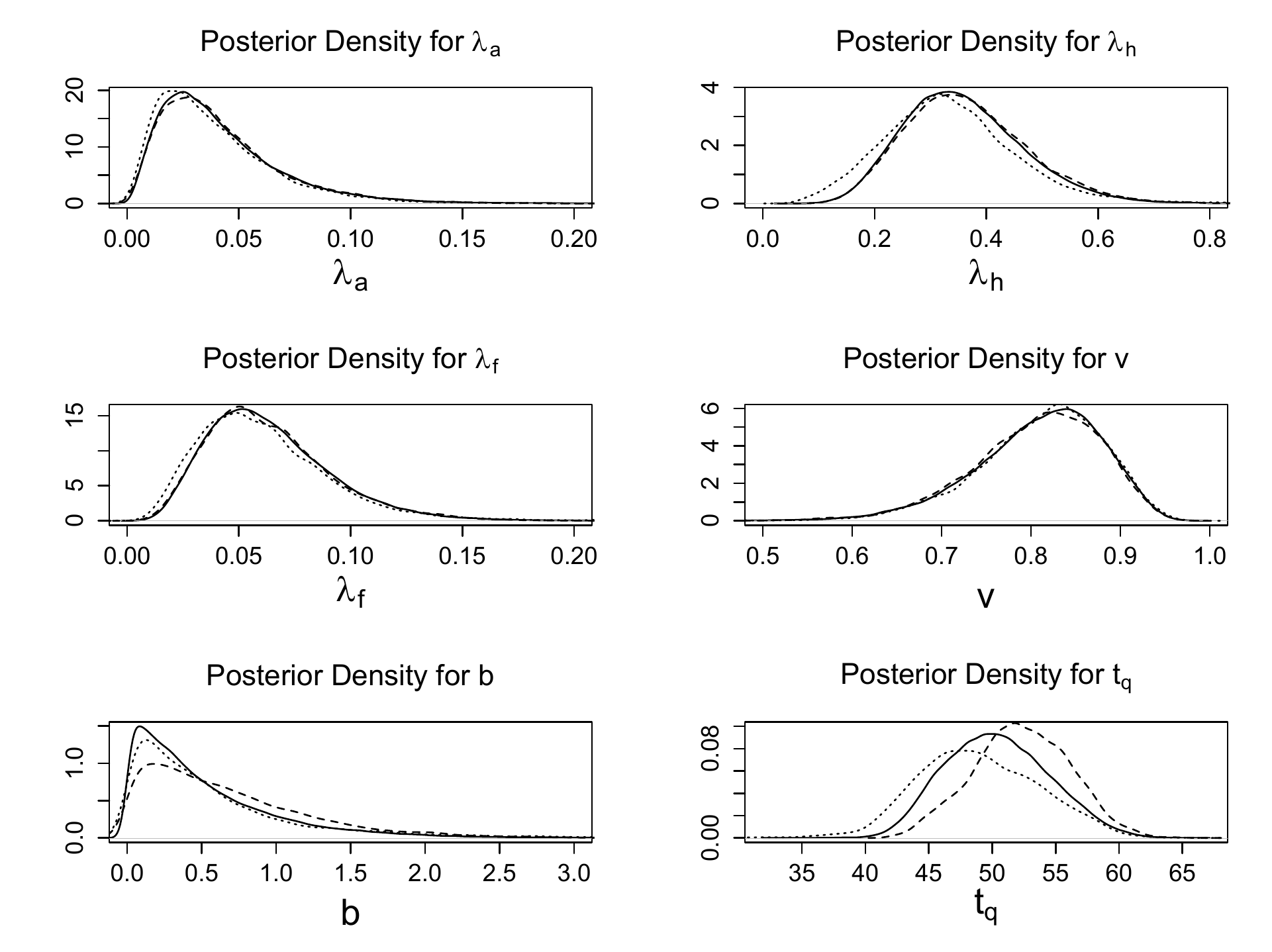}
\includegraphics[width =0.75\linewidth]{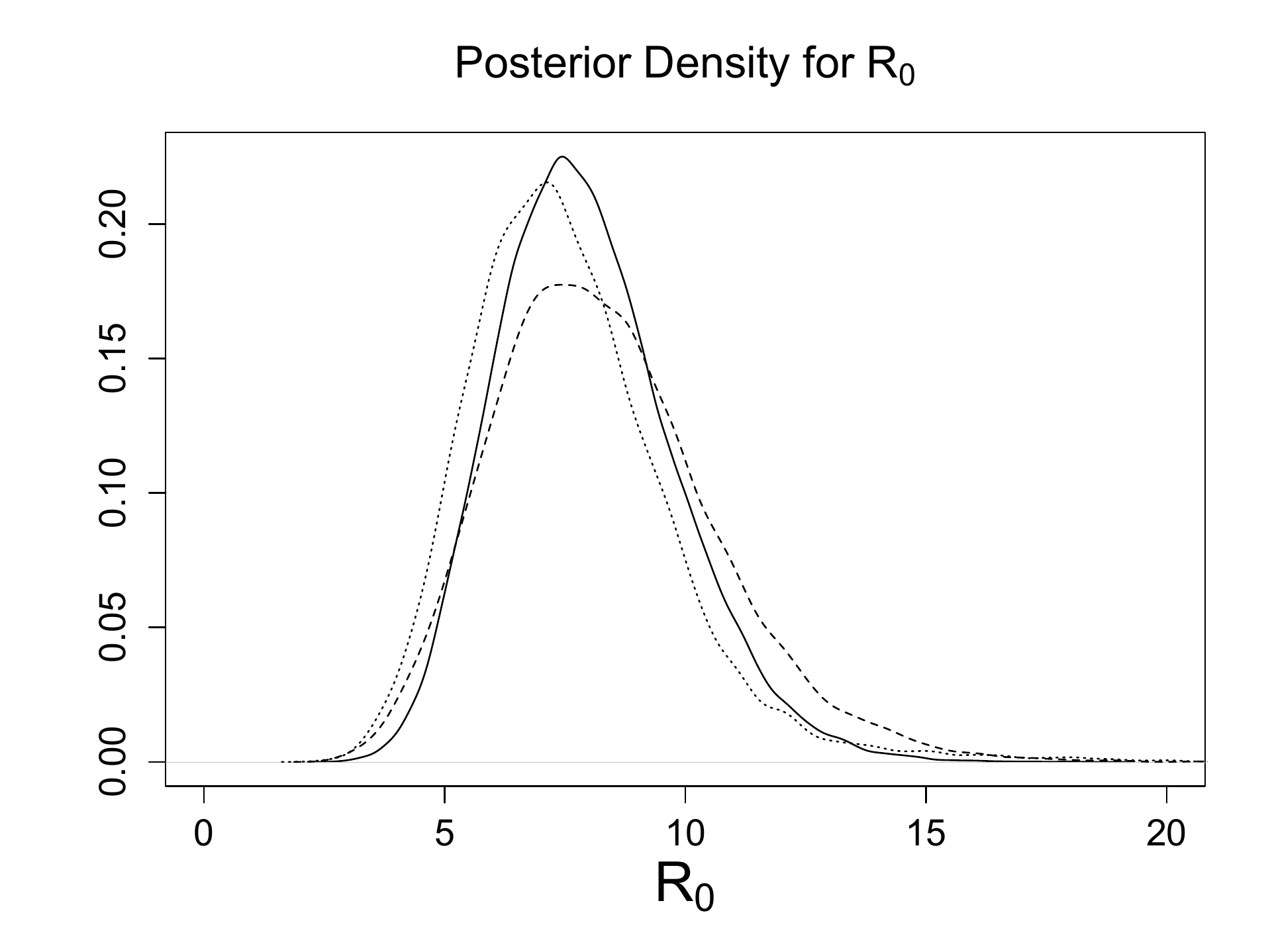}
\includegraphics[scale=0.5, trim={2cm 6cm 4cm 6cm},clip]{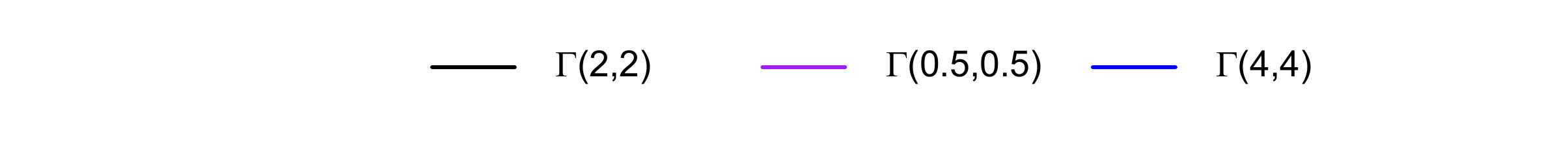}
\caption{Posterior density of the model parameters and $R_0$ as the time taken to quarantine a detected case is varied.  }
\label{fig:gdensity}
\end{figure}

We now briefly explore the sensitivity of our results to the underlying model assumptions, and in particular the assumed values for the periods of time spent in each disease
state. Figure \ref{fig:sdensity} displays posterior densities for the parameters of interest over a range of values taken from \cite{Melt} and \cite{G+L}.
As might be expected, when $\mu_R$, the mean length of the rash period before removal, is reduced to make shorter average infectious periods, the estimates of the infection rate
parameters increase to compensate. It is of interest to note that estimation of $R_0$ is somewhat sensitive to the choice of $\mu_R$. This is likely to be an
artefact of the fact that that there are relatively few cases, the population structure and the introduction of control measures, since in a large uninterrupted outbreak we
would expect $R_0$ to be more or less determined by the outbreak size.
Figure \ref{fig:gdensity} shows the effect of varying $\mu_Q$ and $\sigma_Q$, the parameters which govern the time taken for a case to be quarantined. Specifically, we consider
the effect of halving or doubling the mean time, while keeping the coefficient of variation fixed at unity. Here we see relatively
little impact, which is reassuring since we have very little information on which to base our modelling assumptions, in contrast to those which depend on more generic features
of smallpox.

\subsection{Model assessment}
In order to assess how well our model fits the data we use its posterior predictive distribution. Specifically, we take samples of the basic model parameters
from the MCMC output and simulate the model forwards in time for each set of parameter values. We then compare various aspects of the observed data to the
ensemble of simulations to see if the former aligns in some sense with the latter.

\begin{figure}
\centering
\includegraphics[scale=0.75]{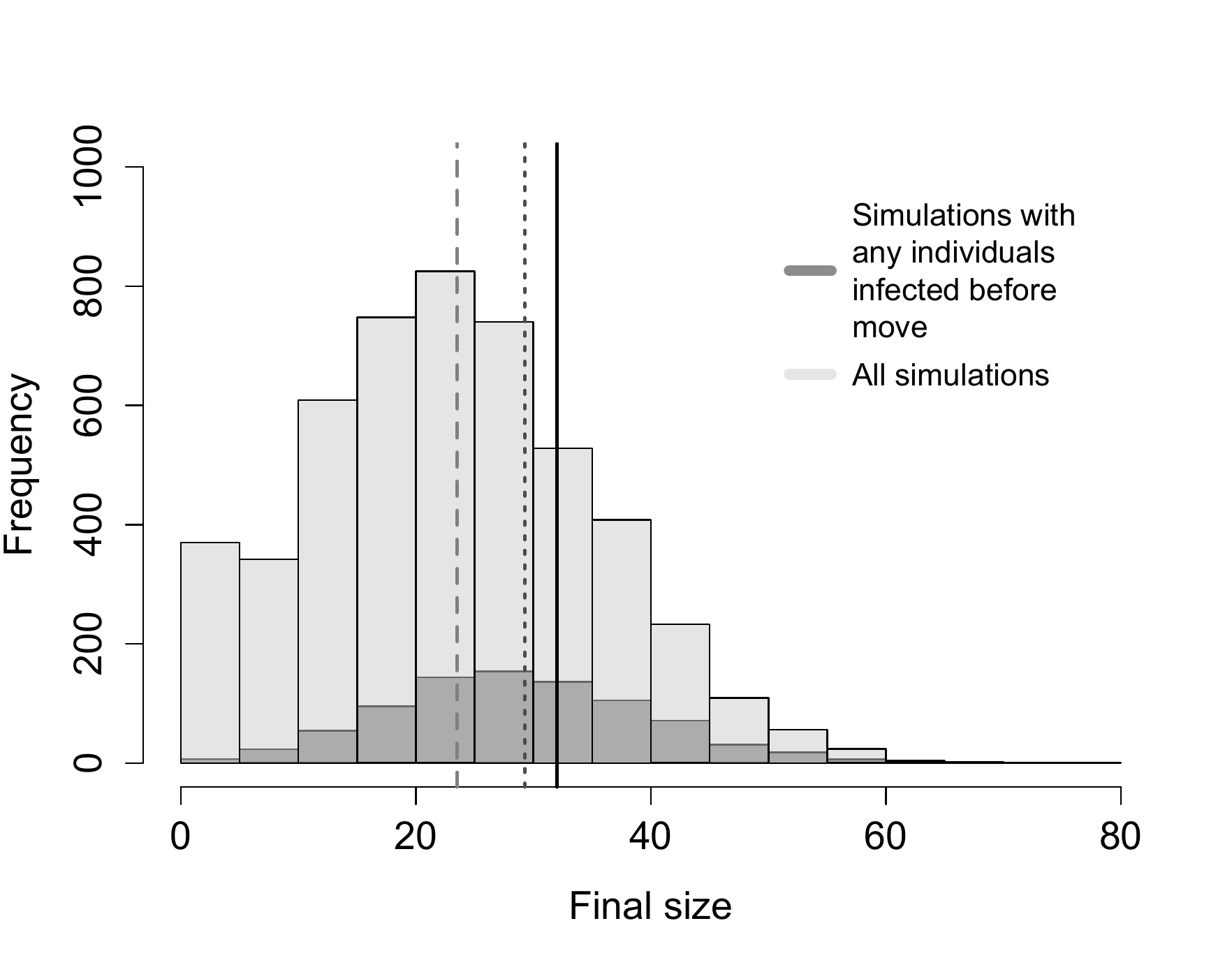}
\caption{Final size of 5000 outbreaks simulated from posterior estimates (mean 23.5, dashed line), and the subset of 848 outbreaks in which at least one of the four individuals
who moved compound was infected prior to the move (mean 29.3, dotted line). The solid line shows the observed final size (32).}
\label{fig:finalsize}
\end{figure}

We start with the final size of the epidemic, i.e. the total number of cases. Figure \ref{fig:finalsize} shows that although the observed final size (32) is not
untypical of those produced by the model, it is some way from the mean (23.5) and mode of the distribution. This underestimation appears to be largely due to the fact
that in the actual outbreak, four individuals, of whom two were infected, moved from compound 1 to compound 2, leading to new cases in compound 2. To account
for this, Figure \ref{fig:finalsize} also shows a histogram of the final size distribution among those simulated epidemics in which at least one of the four
moving individuals was infected. It can be seen that this adjustment gives a better fit to the observed final size.

\begin{figure}
\centering
\includegraphics[scale=0.75]{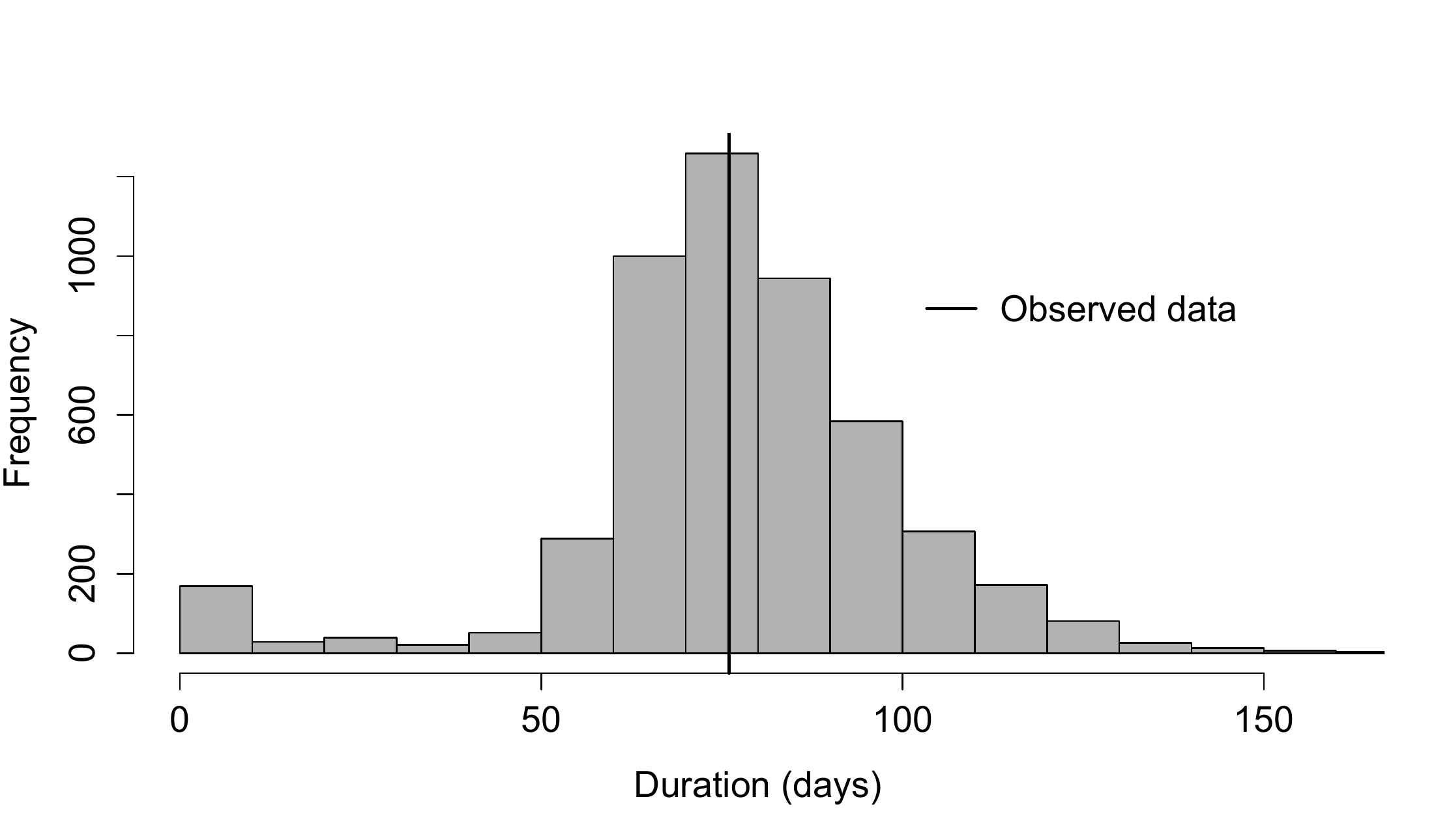}
\caption{Duration of 5000 outbreaks simulated from posterior estimates, with the solid line showing the duration of the Abakaliki outbreak.}
\label{fig:duration}
\end{figure}

We next consider epidemic duration, defined as the length of time between the first case detection (rash) time and the last. Figure \ref{fig:duration} shows a histogram
of the durations of 5000 simulated outbreaks. The mean duration is 76.8 days, which is very similar to the Abakaliki outbreak (76 days). Including only those outbreaks in which
infected individuals moved compound only increases the mean by a few days.

\begin{figure}
\centering
\includegraphics[width=1.00\textwidth]{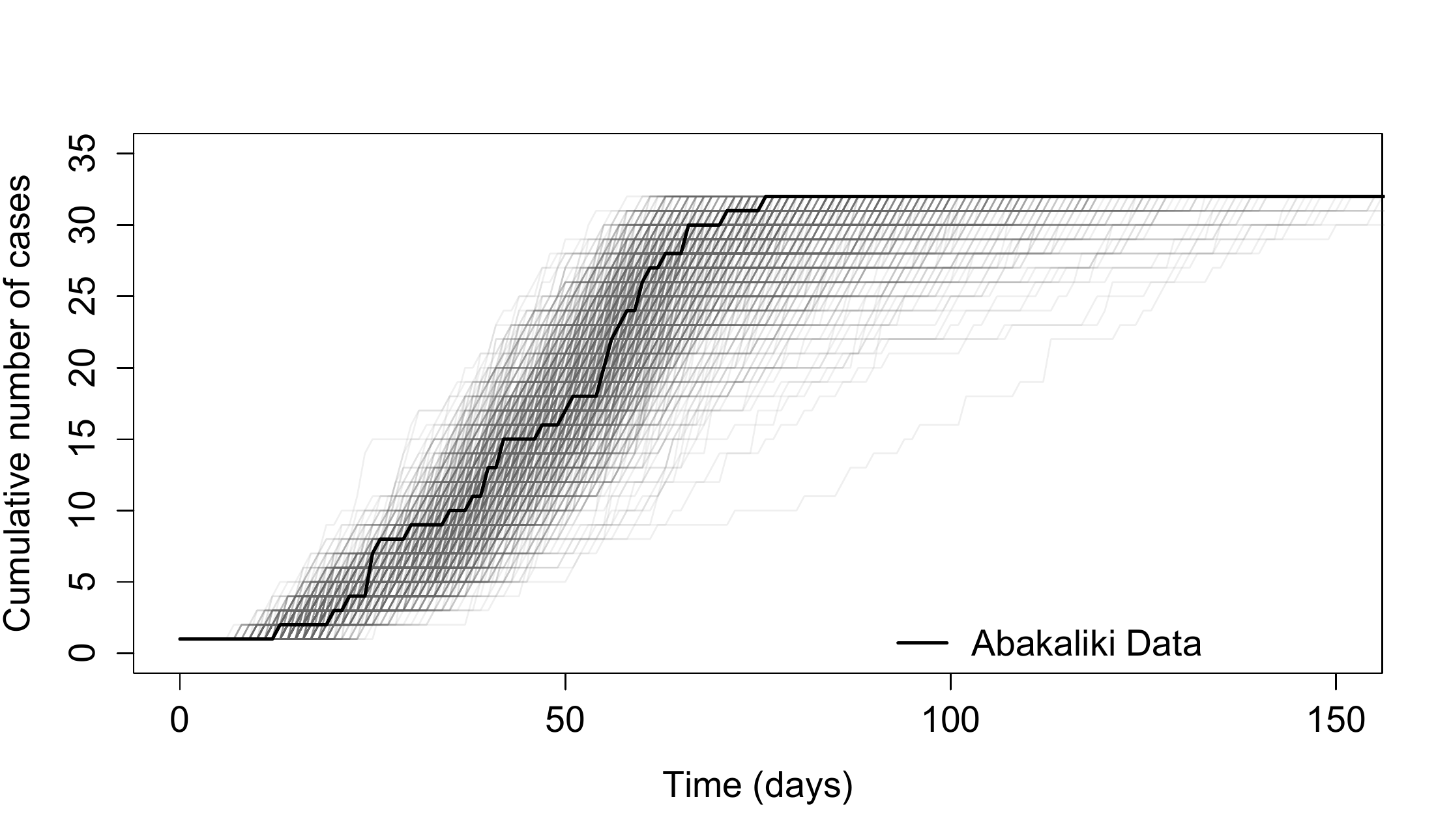}
\caption{1000 simulated epidemics each with a total of 32 cases. The solid line is the observed cumulative number of cases.}
\label{fig:incidence}
\end{figure}

We next compare the simulated cumulative number of cases with the Abakaliki data. This is complicated by the fact that different simulated outbreaks
usually have different total numbers of cases, and so to facilitate the comparison we consider only simulated outbreaks that have the same total
number of cases (32) as the data. Figure \ref{fig:incidence} shows the results of 1000 simulations, from which it appears that the observed data are
reasonably well captured by the model behaviour. To quantify this more precisely, we calculated a posterior predictive p-value as follows. Recall the
chi-squared discrepancy measure (see e.g. \cite{Gelman}), which here takes the form
\begin{equation*}
D(\mathbf{r}, \mathbf{\Phi}) = \sum_{j} \frac{\big(r_j - \mathbb{E}(r_j | \mathbf{\Phi})\big)^2}{\text{Var}(r_j | \mathbf{\Phi})},
\end{equation*}
where $\mathbf{\Phi}$ denotes the model parameters and $\mathbf{r} = (r_1, \ldots, r_{32})$ denotes a vector of case-detection (rash) times.
Note that neither the mean nor variance term are available analytically, and so in practice they are obtained via simulation: given $\mathbf{\Phi}$,
we simulate epidemics repeatedly until we have a sample of size $M_1$, all with 32 cases. The mean and variance of the $j$th rash time is then estimated
directly from this sample.
Suppose now that we have $M$ samples from the posterior density of the model parameters, labelled $\mathbf{\Phi}^{(1)}, \ldots, \mathbf{\Phi}^{(M)}$.
We use the $i$th sample to obtain a simulated epidemic with 32 cases and rash times $\mathbf{r}^{\text{rep}}_i$ by repeatedly simulating epidemics until we
obtain one with 32 cases. Denoting a typical simulation replicate $\mathbf{r}^{\text{rep}}$ and letting $\mathbf{r}^{\text{obs}}$ denote the observed rash times,
the posterior predictive p-value is defined as
\begin{eqnarray*}
\text{ppp-value} & = & \mathbb{P}\Big(D(\mathbf{r}^{\text{rep}},\mathbf{\Phi}) \geq D(\mathbf{r}^{\text{obs}},\mathbf{\Phi}) | \mathbf{r}^{\text{obs}}\Big) \\
& \approx & \frac{1}{M}\sum_{i=1}^{M} \mathbbm{1}_{\{D(r^{\text{rep}}_i,\mathbf{\Phi}^{(i)}) \geq D(\mathbf{r}^{\text{obs}},\mathbf{\Phi}^{(i)})\}}.
\end{eqnarray*}

To interpret this quantity, note that if typical simulations are close to the observed data then we would expect the ppp-value to be around 0.5, while values close
to 0 or 1 would indicate a poor model fit. We carried out this procedure with $M_1 = M = 100$ and obtained a value of 0.42, which is suggestive of a good model fit.
A more accurate value could in principle be obtained using larger values of $M$ and $M_1$, but the procedure is highly time-consuming in practice due to the fact that we require
simulated epidemics of a given final size.

As a final assessment of model fit, from 5000 simulations we found that on average 91\% of those infected were FTC members, compared with 94\% in
the data, although we also found that around 20\% of those infected were from outside the compounds, compared to no such individuals in the data.


\section{Discussion}\label{sec:fut}

{\em Transmission} The posterior estimates of the basic model parameters indicate clearly that the dominant mode of transmission was within-compound between individuals of
the same confession. Our estimates of
who-infected-whom align with this; for instance, from Figure \ref{whoinfwhom}, the vast majority of transmission events occurred within a compound in the most-likely infection pathway. The epidemiological investigation reported in \cite{thompsonfoege68} found that spread within compounds, and within families in particular, appeared to drive the epidemic and that membership of FTC itself was not the primary transmission route. This is in agreement with our findings. As in \cite{E+D}, we found that infectiousness is markedly less during the fever
period than the rash period, although our mean estimate of the reduction parameter $b$ is larger than their maximum likelihood value, which is most likely due to the skewed shape of
the marginal posterior density.

{\em Reproduction numbers} Our posterior mean estimate of $R_0$ is close to 8. This is slightly larger than the Eichner and Dietz estimate (6.87) but underlines the
potentially devastating nature of smallpox. Such values are radically different from those obtained using simpler models: for instance, assuming an SIR model in a
homogeneously-mixing population of 120 FTC individuals typically results in $R_0$ estimates slightly larger than one, even allowing for the infection rate to vary
with time (see e.g. \cite{ON+R}, \cite{Xu16}). This highlights the importance of models which properly take population structure into account. As previously stated, $R_0$ can be interpreted as the average number of secondary cases produced by a single infective individual in a large susceptible population. For the Abakaliki data, such an interpretation is hard to apply directly since the compounds, wherein most transmission occurs, are small enough to provide a rapid saturation effect via the depletion of available susceptible individuals.

{\em Model adequacy} Our model appears to fit the data reasonably well, with the possible caveat that the model invariably predicts cases occurring outside of the compounds. The fact that the entire population of Abakaliki is rather unrealistically modelled as a homogeneously mixing population goes some way to explaining this; in particular, the potential for
contacts between those inside and outside the compounds, and especially between FTC members and those outside the compounds, could well have been rather less than that assumed
in the model. According to \cite{thompsonfoege68}, the FTC community was largely isolated from the community at large, although several of its adult members were involved in trading activities in and around Abakaliki. Consequently, a model in which some fraction of FTC members had contact with the outside community might be more realistic, although there are
no data to directly inform this. Another aspect that is missing from our model is that of age categories;  \cite{thompsonfoege68} state that the highest attack rates were among
children. However, there do not appear to be sufficient data on compound composition to accurately incorporate age categories, and it seems likely that a model with age-specific
transmission rates may be over-parameterised.

{\em Control measures and the end of the outbreak} It seems likely that the advent of control measures at time $t_q$ played a crucial role in bringing the outbreak to its conclusion
rapdily. Under the model assumptions, control measures reduce the rash period from an average of 16 days to just 2 days, which in turn reduces the number of new infections. Interestingly, the posterior mean of $R_0$ after $t_q$ (i.e. $R_0$ with $\mu_R = \mu_Q = 2.0$) is around 1.5, but this in itself is insufficient to permit further large-scale
spread due to the depletion of susceptibles within the compounds, and the fact that the epidemic in the population outside the compounds is sub-critical (i.e. the basic reproduction
number is less than 1). Expanding the latter point, if we define pre- and post-control measure reproduction numbers for spread within compounds, FTC and the wider population (e.g. $R_a = (\mu_R + b \mu_F)\lambda_a$, etc.), then posterior mean estimates show that (i) within compounds, the epidemic is super-critical before and after $t_q$; (ii) with the FTC community, the epidemic switches from super- to sub-critical; (iii) in the wider population, the epidemic is always sub-critical. Despite this, increasing the value of $t_q$ in simulations was found to
increase the outbreak size; for instance, setting $t_q$ to be 50, 100 and 200 gave mean outbreak sizes of around 24, 44 and 64, respectively. However, with no restrictions, we found
the average outbreak size to be around 86, which underlines the fact that the epidemic was sub-critical in the wider population.

{\em Accuracy of the Eichner and Dietz likelihood approximation}
It is of interest to see that our results are fairly similar to those obtained by \cite{E+D}. The most plausible explanation for this is the fact that distributions used
for the length of time in each disease stage do not have particularly large variances, which in turn means that the model is not all that different to one in which all event
times are assumed known. For such a model, the approximation method used by Eichner and Dietz gives the true likelihood, essentially because the distributions used to
approximate uncertain event times collapse to point masses around the true values. A further point of interest is that the Eichner and Dietz approximation produces a likelihood
function which is numerically but not analytically tractable, specifically because it involves integrals that must be evaluated numerically. Although this is sufficient for
optimization purposes such as maximum likelihood, in practice such likelihood functions can be computationally prohibitive for use within MCMC algorithms since they must be
repeatedly evaluated. It would therefore be of interest to develop analytically tractable approximate likelihood functions.


\section{Acknowledgements}
It is a pleasure to thank Martin Eichner for helpful discussions about this work.
J.E.S. was funded by a UK Engineering and Physical Sciences Research Council studentship.




\end{document}